\documentclass[preprint,pra,aps,showpacs,preprintnumbers,amsmath,amssymb,eqsecnum]{revtex4}
\usepackage{graphicx}% Include figure files
\usepackage{dcolumn}% Align table columns on decimal point
\usepackage{bm}% bold math

\def\x{{\bf x}}

\def\k{\mathbf{k}}
\def\q{\mathbf{q}}
\def\a{\alpha}
\def\b{\beta}
\def\la{\langle}
\def\ra{\rangle}
\def\ria{\rightarrow}
\def\a{\alpha}
\def\au{{\underline{\alpha}}}
\def\s{{\sigma}}
\def\half{\frac{1}{2}}

\newcommand\beq{\begin{equation}}
\newcommand\eeq{\end{equation}}
\newcommand\bea{\begin{eqnarray}}
\newcommand\eea {\end{eqnarray}}

\begin{document}

%\preprint{APS/123-QED}

\title{Macroscopic Superpositions, Decoherent Histories and
the Emergence of Hydrodynamic Behaviour}

\author{J.J.Halliwell}%
\affiliation{Blackett Laboratory \\ Imperial College \\ London SW7
2BZ \\ UK }

%\affiliation{

\date{\today}% It is always \today, today,

\begin{abstract}

Macroscopic systems are described most completely by local densities
(particle number, momentum and energy) yet the superposition states
of such physical variables, indicated by the Everett interpretation,
are not observed. In order to explain this,
it is argued that histories of local number, momentum and energy
density are approximately decoherent when coarse-grained over
sufficiently large volumes. Decoherence arises directly from the
proximity of these variables to exactly conserved quantities
(which are exactly decoherent), and not from
environmentally-induced decoherence. We discuss the approach to
local equilibrium and the subsequent emergence of hydrodynamic
equations for the local densities. The results are general but we
focus on a chain of oscillators as a specific example in which
explicit calculations may be carried out. We discuss the relationships
between environmentally-induced and conservation-induced decoherence and
present a unified view of these two mechanisms.

\vskip 1.0in

%\centerline{Imperial/TP/2-03/23, May 2003}

\end{abstract}

%\pacs{03.65.-w, 03.65.Yz, 03.65.Ta, 05.70.Ln}% PACS, the Physics and Astronomy
\maketitle

\section{Introduction}

If the Everett interpretation of quantum theory is to be taken
seriously, there will exist superposition states for macroscopic
systems, perhaps even for the entire universe. Since such states
are not observed, it is therefore necessary to explain why they go
away. This question is a key part of the general question of the
emergence of classical behaviour from quantum theory, an issue
that has received a considerable amount of attention \cite{Har6,Harvol}.

There are a number of different
approaches to emergent classicality, but common to most of them is the notion that
there must be decoherence, that is, that certain types of quantum
states of the system in question exhibit negligible interference,
and therefore superpositions of them are effectively equivalent to
statistical mixtures.
Decoherence has been extensively investigated for the situation
in which there is a distinguished system, such as a particle,
coupled to its surrounding environment \cite{JoZ,Zur}.
However, for many macroscopic systems, and in particular for
the universe as a whole, there may be no natural split
into distinguished subsystems and the rest and another way of
identifying the naturally decoherent variables is required.
Most generally, decoherence comes about when the
variables describing the entire system of interest naturally
separate into ``slow'' and ``fast'', whether or not this
separation corresponds to, respectively, system and environment.
If the
system consists of a large collection of interacting identical
particles, such as a fluid for example, the natural set of slow
variables are the local densities: energy, momentum, number,
charge {\it etc.} They are ``slow'' because they are locally
conserved. These variables, in fact, are also the variables
which provide the most complete description of the classical state
of a fluid at a macroscopic level. The most general demonstration
of emergent classicality therefore consists of showing that, for a
large collection of interacting particles described
microscopically by quantum theory, the local densities become
effectively classical. Although decoherence through the
system--environment mechanism may play a role, since the
collection of particles are coupled to each other, it is important
to explore the possibility that, at least in some regimes,
decoherence could come about
because the local densities are almost conserved if averaged over
a sufficiently large volume \cite{GH2}. Hence, the approximate
decoherence of local densities would then be due to their proximity
to a set of exactly conserved quantities, and
exactly conserved quantities obey superselection rules.

We will approach these questions using the decoherent histories
approach to quantum theory \cite{GH2,GH1,Gri,Omn,Hal1,Hal5}. This approach
has proved particularly useful for discussing emergent
classicality in a variety of contexts. In particular the
issues outlined above are most clearly expressed in the language
of decoherent histories. The central object of interest is the
decoherence functional,
\begin{equation}
D (\au, \au') = {\rm Tr} \left( P_{\a_n} (t_n) \cdots P_{\a_1}
(t_1) \rho P_{\a_1'} (t_1) \cdots
P_{\a_n'} (t_n) \right)
\label{1.1}
\end{equation}
The histories are characterized by the initial state $ \rho
$ and by the strings of projection operators $P_{\a} (t)$
(in the Heisenberg picture) at times
$t_1$ to $t_n$ (and $\au$ denotes the string of alternatives $\a_1
\cdots \a_n$). Intuitively, the decoherence functional is a
measure of the interference between pairs of histories $\au$,
$\au'$. When it is zero for $\au \ne \au' $, we say that the
histories are decoherent and probabilities  $ p (\au ) = D (\au,
\au ) $ obeying the usual probability sum rules may be assigned to
them. One can then ask whether these probabilities are strongly
peaked about trajectories obeying classical equations of motion.
For the local densities, we expect that these equations will be
hydrodynamic equations.

The aim of this paper is review this programme, following primarily
Refs.\cite{Hal10,Hal2,Hal3,Hal4}. We will outline
the argument showing how the approximate conservation of
the local densities implies negligible interference of their histories
at sufficiently coarse grained scales, and show how hydrodynamic equations
of motion for them arise.

\section{Local Densities and Hydrodynamic Equations}

We are generally concerned with a system of $N$ particles
described at the
microscopic level by a Hamiltonian of the form
\begin{equation}
H = \sum_j
{ {\bf p}_j^2 \over 2 m } + \sum_{\ell >j } V_{j\ell} (
\q_j - \q_\ell )
\end{equation}
We are particularly interested in the number density $n(\x)$, the
momentum density ${\bf g}(\x)$ and the energy density $ h (\x )$,
defined by,
\begin{eqnarray}
n(\x) &=& \sum_j \ \delta(\x -\q_j)
\\
{\bf g}(\x) &=& \sum_j \ {\bf p}_j \ \delta (\x- \q_j)
\\
h(\x) &=& \sum_j \  { {\bf p}^2_j \over 2 m } \delta(\x- \q_j) + \sum_{\ell >
j} V_{j \ell} (\q_j - \q_{\ell} ) \delta(\x- \q_j)
\end{eqnarray}
We are interested in the integrals of these quantities over
volumes which are large compared to the microscopic scale but
small compared to macroscopic physics.
Integrated over an infinite volume, these become the total particle
number $N$, total momentum $P$ and total energy $H$, which are exactly
conserved.
It is also often more useful to work with the Fourier transforms of the
local densities,
\begin{eqnarray}
n(\k) &=& \sum_j \ e^{i \k \cdot \q_j}
\label{6}
\\
{\bf g}(\k) &=& \sum_j \ {\bf p}_j \ e^{i \k \cdot \q_j}
\label{7}
\\
h(\k) &=& \sum_j \  { {\bf p}^2_j \over 2 m} e^{i \k \cdot \q_j}+
\sum_{\ell > j} V_{j \ell} ( \q_j - \q_{\ell} ) \ e^{i \k \cdot
\q_j}
\label{8}
\end{eqnarray}
These quantities tend to the exactly conserved quantities in the
limit $ k = | \k | \ria  0 $, so we are interested in what happens
in what happens for small but non-zero $k$.

Setting aside for the moment the issues of decoherence,
there is a standard technique for deriving hydrodynamic equations
for the local densities \cite{Harvol,Hua,hydro}. It starts with the continuity equations
expressing local conservation, which have the form,
\begin{equation}
\frac{ \partial \sigma} { \partial t} + \nabla \cdot {\bf  j } = 0
\label{1.9}
\end{equation}
where $\sigma $ denotes $n$, ${\bf g}$ or $h$ (and the current
${\bf j}$ is a second rank tensor in the case of ${\bf g}$).
It is then assumed that, for a wide variety of initial states,
conditions of local equilibrium are
established after a short period of time.
This means that on scales small compared to the overall
size of the fluid, but large compared to the microscopic scale,
equilibrium conditions are reached in each local region,
characterized by a local temperature, pressure {\it etc.} which vary
slowly in space and time. Local equilibrium is described by
the density operator
\begin{equation}
\rho = Z^{-1} \exp \left( - \int d^3x \ \b (\x) \left[ h(\x) - \bar \mu (\x)
n (\x) - {\bf v} (\x) \cdot {\bf g} (\x) \right] \right)
\label{1.10}
\end{equation}
where $ \b  $, $\bar \mu $ and $ {\bf v} $ are Lagrange multipliers and
are slowly varying functions of space and time. $\b$ is the inverse
temperature, ${\bf v}$ is the average velocity field, and $\bar \mu$
is related to the chemical potential which in turn is related to
the average number density. (Note that the local equilibrium state is defined in
relation to a particular coarse-graining, here, the anticipated calculation
of average values of the local densities. Hence it embraces all possible states
that are effectively equivalent to the state Eq.(\ref{1.10}) for the purposes of calculating
those averages.) The hydrodynamic
equations follow when the continuity equations are averaged in this
state. These equations form a closed set because the local equilibrium
form depends (in three dimensions) only on the five Lagrange multiplier
fields $ \b, \bar \mu $, $ {\bf v}$ and there are exactly five continuity equations
(\ref{1.9}) for them. (More generally, it is possible to have closure up to a set
of small terms which may be treated as a stochastic process. See Refs.\cite{BrHa,CaH1},
for example.)

We will in this paper concentrate on the useful pedagogical
example of a chain of oscillators, in which many calculations can be
carried out explicitly \cite{Hal10}. The Hamiltonian of this system is
\begin{equation}
H = \sum_{n=1}^N \left[ { p_n^2 \over 2m} + {\nu^2 \over 2} ( q_n
- q_{n-1} )^2 + { K \over 2} ( q_n - b_n )^2 \right]
\label{Ham}
\end{equation}
There are two cases $K=0$ (the
simple chain) and $K \ne 0 $ (the harmonically bound chain). In
the bound chain case, it is also useful to consider the case $b_n
= 0$, which corresponds to the situation in which the whole chain
moves in a harmonic potential. We
consider a finite number $N$ of particles but it is sometimes
useful to approximate $N$ as infinite.

The local densities of this system are
\begin{eqnarray}
n(x) &=&  \sum_{n=1}^N \delta (q_n - x )
\label{5.1}\\
g(x) &=& \sum_{n=1}^N p_n \delta (q_n - x )
\label{5.2}\\
h(x) &=& \sum_{n=1}^N \left[ { p_n^2 \over 2m} + {\nu^2 \over 2} (
q_n - q_{n-1} )^2 + \half K ( q_n - b_n )^2 \right] \delta (q_n -
x ) \label{5.3}
\end{eqnarray}
They satisfy the local conservation laws
\begin{eqnarray}
\dot n(x) &=& - \frac {1} {m} \frac {\partial g} {\partial x}
\label{6.1} \\
\dot g (x) &=& - \frac {\partial \tau} {\partial x} - K x n(x) + K \sum_j b_j \delta (q_j - x)
\label{6.2} \\
\dot h(x) &=& - \frac {\partial j} {\partial x}
\label{6.3}
\end{eqnarray}
The currents $ \tau (x)$ and $j(x)$ are rather complicated in configuration space, except in the case where
we neglect the interaction term, when they are given by
\begin{eqnarray}
\tau (x) &=& \sum_j \frac {p_j^2} {m} \delta (q_j - x)
\label{6.9} \\
j(x) &=& \sum_j \frac {p_j} {m} \left( \frac {p_j^2} {2m} + \half K (q_j -b_j)^2 \right)
\delta (q_j -x )
\label{6.10}
\end{eqnarray}

%Equations (\ref{6.1})--(\ref{6.3}) do not in general form a closed system, so do not
%lead to hydrodynamic equations. To get a closed set, it is necessary to average
%these equations in a set of states depending on just three fields, thereby
%obtaining three equations for three unknowns. In the standard approach to
%deriving hydrodynamics, the local equilibrium state is chosen. We will discuss this
%below in Section C, but first we consider the simpler and instructive case
%of the normal mode coherent states.

The standard derivation of the hydrodynamic equations may be carried out reasonably
easily in this model. Instead of the density operator form Eq.(\ref{1.10}) of the local
equilibrium state, we work with the equivalent one-particle Wigner function (phase space
density)
\begin{equation}
w_j (p_j,q_j) = f(q_j) \exp \left( - \frac { (p_j- m v(q_j))^2 } {2 m k T(q_j)} \right)
\label{6.11}
\end{equation}
where $f$, $v$ and $T$ are slowly varying functions of space and time ($f$ is simply related
to the chemical potential in Eq.(\ref{1.10})). This is the one-particle
distribution function for particle $j$ -- it is labelled by $j$ since the particles are distinguishable.
If we now average
the system Eqs.(\ref{6.1})--(\ref{6.3}), together with the currents $\tau (x)$, $j(x)$
in the local equilibrium state, we obtain a closed
system, since we get three equations for three unknowns. In the case of negligible interactions
and $b_j = 0$, we find
\begin{eqnarray}
\la n(x) \ra &=& N f(x)
\label{6.12} \\
\la  g (x) \ra &=& m v(x) N f(x)
\label{6.13} \\
\la h(x) \ra &=& \left(  \half m v^2 + \half k T + \half K x^2 \right) N f (x)
\label{6.14} \\
\la \tau (x) \ra &=&  \left( mv^2 + k T \right) N f(x)
\label{6.15} \\
\la j(x) \ra &=& \left( \frac {3} {2} v k T  + \half m v^3 \right) N f(x) + \frac {K} {2m} x^2 \la g(x) \ra
\label{6.16}
\end{eqnarray}
The first three equations give the explicit inversion between the averages of the local densities
and the three slowly varying functions $f,v,T$.
Inserted in Eqs.(\ref{6.1})--(\ref{6.3}), the above relations give
a closed set of equations for the three variables $f$, $v$ and
$T$. After some rearrangement, these equations are
\begin{eqnarray}
\frac {\partial f} {\partial t} + v \frac {\partial f} {\partial x}
&=& - f \frac {\partial v} {\partial x}
\\
\frac {\partial v} {\partial t} + v \frac {\partial v} {\partial x}
&=& - \frac {1} {m } \frac {\partial \theta } {\partial x}
-\frac {\theta} {m f} \frac {\partial f } { \partial x}
- \frac {K x } {m}
\\
\frac {\partial \theta } {\partial t} + v \frac {\partial \theta } {\partial x}
&=& - 2 \theta \frac {\partial v} {\partial x}
\end{eqnarray}
where $\theta = k T $. These are the equations for a one-dimensional fluid moving
in a harmonic potential \cite{Hua}. Note that non-trivial equations are obtained even
though we have neglected the interaction terms in deriving them. The role
of interactions is to ensure the approach to local equilibrium,
as we discuss below.

In these expressions, the definition of
the temperature fields is essentially equivalent to,
\begin{equation}
\sum_j \frac {1} {2m} ( \Delta p_j)^2  \ \delta (q_j - x) = \half
k T(x) n (x) \label{6.18}
\end{equation}
(recalling that we are working at long wavelengths, so the
$\delta$-function is coarse-grained over a scale of order
$k^{-1})$. Hence temperature arises not from an environment, but
from the momentum fluctuations averaged over a coarse-graining
volume.

It is straightforward to give a decoherent histories version
of the standard derivation of the hydrodynamic equations.
We take the initial state to be the local equilibrium state.
We take the histories to be characterized by projection
operators onto broad ranges of values of the local densities.
(The local densities do not commute in general, but they will
approximately commute for sufficiently small $k$ and it is
not difficult to construct quasi-projectors that are well-localized
in all three densities). Then, it is easily shown that for sufficiently
broad projections, the histories are peaked about the average values
of the local densities, averaged in the initial local equilibrium
state \cite{Hal3}. The standard derivation shows that the average values obey hydrodynamic
equations hence the probabilities are peaked about evolution according to
those equations.

However, what is important here is that
the decoherent histories approach to quantum theory offers the possibility
of a derivation of emergent classicality much more general than that
entailed in the standard derivation of hydrodynamics. The standard
derivation is rather akin to the Ehrenfest theorem of elementary quantum
mechanics which shows that the averages of position and momentum operators obey
classical equations of motion. Yet a description of emergent classicality
must involve much more than that \cite{Har6}.
Firstly, it must demonstrate decoherence of the local
densities, thereby allowing us to talk about probabilities for their histories.
Secondly, it should not be restricted to a special initial state.
Whilst it is certainly plausible that many initial states will tend to
the local equilibrium state, the standard derivation
does not obviously apply to superpositions of macroscopic states, which are exactly the states
a description of emergent classicality is supposed to deal with. It is to this more general
derivation that we now turn.

\section{Decoherence and Conservation}

We begin by describing the connection between decoherence and conservation.
It is well-known that histories of exactly conserved quantities are exactly
decoherent \cite{HLM}. The simple reason for this is that the projectors
onto conserved quantities commute with the Hamiltonian. The projectors $
P_{\a_k}$ on one side of the decoherence functional (\ref{1.1}) may therefore
be brought up against the projectors $P_{\a_k'} $ on the other
side, hence the decoherence functional is exactly diagonal.
(In the situation considered here, in which there are three conserved
quantities involved, these quantities must in addition commute
with each other, but this is clearly the case.)

There is another way of expressing this that is more useful for the
generalization to the local densities. Suppose we take the initial
state to be a pure state $ | E, {\bf P}, N \rangle $ which is an eigenstate
of the total energy, momentum and number, and consider a history of
projections onto total energy, momentum and number. Clearly, unitary evolution
preserves the eigenstate (except for a phase), and the projections acting on
it either give back the state, or give zero. This means that
\beq
P_{\a_n} (t_n) \cdots P_{\a_1} (t_1) | E, {\bf P}, N \rangle
\eeq
is equal to either $ | E, {\bf P}, N \rangle $ or to zero.
It is easy to see, by expanding
an arbitrary initial state in eigenstates of the conserved quantities that this implies
exact decoherence of histories for any initial state.

Turning now to the local densities, which are most usefully discussed in
the Fourier transformed form Eqs.(\ref{6})-(\ref{8}), the above argument shows that
they define exactly decoherent sets of histories for the case $ k = 0$.
Now here is the key point: as $k$ departs from zero, the decoherence functional
will depart from exact diagonality, but there will still clearly be approximate
decoherence if $k$ is sufficiently small. That is, decoherence of local densities
essentially follows from an expansion for small $k$ about the exactly decoherent
case, $k=0$. The aim of much of the rest of
this section is to spell out in more detail how this works out.

We generalize the above argument for exact decoherence of histories
of conserved quantities, to locally conserved quantities. We suppose we have
a set of histories characterized by projections onto the local densities
for some value of $k$.
We then consider
states $ | h, {\bf g}, n \rangle $ which are approximate eigenstates
of the local densities. Exact eigenstates are not possible,
but it is not hard to find states which are well-localized all three
variables.
Under time evolution, the local density eigenstates $  | h, {\bf g}, n \rangle $
will not remain exact eigenstates, but as long as they remain approximate
eigenstates (that is, well-localized in the local densities), the above
argument goes through and we deduce approximate decoherence.

Hence, denoting the local densities
by $Q$, what we need to show is that, for an initial state localized
in the local densities, under time evolution, $Q$ satisfies the condition,
\begin{equation}
{ \left( \Delta Q (t) \right)^2 \over \la Q (t) \ra^2 } << 1
\label{2.2}
\end{equation}
where
\begin{equation}
\left( \Delta Q (t) \right)^2 = \la Q^2(t) \ra - \la Q(t) \ra^2
\end{equation}
Eq.(\ref{2.2}) means that the state remains strongly peaked in the variable $Q$
under time evolution. The states are then approximate eigenstates
of the projectors at each time as long as the widths of the projectors
are chosen to be much greater than $ ( \Delta Q (t) )^2$.
The condition Eq.(\ref{2.2}) must be true approximately for
some $k \ne 0$ since it holds exactly in the limit $k \rightarrow 0 $.
The question is to determine the lengthscale involved.

The number and momentum density are both operators of the form,
\begin{equation}
A = \sum_{n=1}^N A_n
\end{equation}
as is the local energy density, if
we ignore the interaction term. For such operators it follows that
\begin{equation}
( \Delta A)^2 = \sum_n (\Delta A_n)^2 + \sum_{n \ne m} \s
(A_n, A_m)
\label{2.3}
\end{equation} and
\begin{equation}
\la A \ra^2 = \sum_{n,m} \la A_n \ra \la
A_m \ra
\end{equation}
where the correlation function $\sigma$ is defined by
\beq
\sigma (A,B) = \half \langle A B + B A \rangle - \langle A \rangle \langle B \rangle
\eeq
A state will be an approximate eigenstate of the
operator $A$ if
\begin{equation}
\frac { ( \Delta A)^2 } { \la A \ra ^2}  \
\ll \ 1
\end{equation}
The expression for $ \la A \ra^2 $ potentially
involves $N^2$ terms, as does the expression for $ ( \Delta A)^2
$, but the latter will involve only $N$ terms if the correlation
functions $ \s (A_n, A_m ) $ are very small or zero for $n \ne m $.
So simple product states will be approximate eigenstates
and will have $ (\Delta A)^2 / \la  A \ra^2 $
of order $ 1/N$. (See Refs.\cite{Hal2,Hal3} for more detailed examples
this argument).

Under time evolution, the interactions cause
correlations to develop. However, the states will remain
approximate eigenstates as long as the correlations are
sufficiently small that the second term in Eq.(\ref{2.3}) is much
smaller than order $N^2$. The interactions and the subsequent
correlations are clearly necessary in order to get interesting
dynamics and in particular the approach to local equilibrium. The
interesting questions is therefore whether there is a regime
where the effects of interactions are small enough to permit
decoherence but large enough to produce interesting dynamics.
The fact that the variables we are interested in are locally conserved
indicates that there is such a regime.
The important point is that the local
densities become arbitrarily close to exactly conserved quantities as $k \ria 0$.
This means that, at any time, $ ( \Delta A)^2 / \la A \ra^2 $ becomes arbitrarily
close to its initial value (which is of order $1/N$) for sufficiently small $k$.

In specific examples
an uncorrelated initial state develops correlations with a typical lengthscale.
These correlations typically then decay with time. What is found
is that the second term in Eq.(\ref{2.3}) will remain small as long
as $k^{-1}$ is much greater than the correlation length. Hence the key physical
aspect is the locality of the interactions, meaning that only limited local correlations
develop, together with the coarse-graining scale $k^{-1}$ which may be chosen
to be sufficiently large that the correlation scale is not seen.
Differently put, as $k$ increases from zero, departing from exact decoherence,
it introduces a lengthscale $k^{-1}$. Since the decoherence functional is a dimensionless
quantity, clearly nothing significant can happen until $k^{-1}$ becomes comparable
with another lengthscale in the system. The natural scale is the correlation
length in the local density eigenstates.

\section{Chains of Oscillators}

Some of the claims of the physical ideas of the previous
section may be seen explicitly in the oscillator model with Hamiltonian
Eq.(\ref{Ham}).
The equations of motion are
\begin{equation}
m \ddot q_n + K (q_n - b_n)  = \nu^2 ( q_{n+1} - 2 q_n + q_{n-1} )
\end{equation}
where we take $ q_{N+1} = q_1 $. This system has been discussed
and solved in many places \cite{HuR1,HuR2,Fey,Thi,Aga,TeS}.
The solution may be written,
\begin{equation}
q_n (t) = b_n + \sum_{r=1}^N \left[ f_{r-n} (t) q_r (0) + {
g_{r-n} (t) \over m \Omega } p_r (0) \right]
\end{equation}
were, $\Omega^2 = ( K + 2 \nu^2) / m $.
For the bound chain, $ K \ne 0 $, it is most useful to work in the
regime in which the interaction between particles is much weaker
than the binding to their origins, so $ \nu^2  << K $. In this
case, the functions $f_r(t)$ and $g_r(t)$ are given by \cite{HuR1},
\begin{equation}
f_r (t) \approx J _r (\gamma \Omega t ) \cos \left( \Omega t - \pi r / 2 \right)
\end{equation}
and
\begin{equation}
g_r (t) \approx J_r ( \gamma \Omega t ) \sin \left( \Omega t - \pi r / 2 \right)
\end{equation}
where $\gamma = \nu^2 / m \Omega ^2 $, so $\gamma << 1 $ and $J_r$
is the Bessel function of order $r$ (and we have used the convenient
approximation of taking $N$ to be infinite).

The general behaviour of the solutions is easily
seen. The functions $f_{r-n}(t)$ and $g_{r-n} (t)$ loosely
represent the manner in which an initial disturbance of particle
$r$ affects particle $n$ after a time $t$, and is given
by the properties of Bessel functions \cite{AbSt}. $J_n(x)$ decays rapidly for large $n$
at fixed $x$, so distant particles do not affect each other very much.
Evolving in $x$, $J_n(x)$ starts
at zero for $x=0$ (except for $n=0$, where $J_0 (0) = 1$),
grows exponentially, and then goes into a slowly decaying oscillation,
\begin{equation}
J_n (x) \sim \left( \frac {2} {\pi x} \right)^{1/2} \ \cos \left(
x - \pi n / 2 - \pi/ 4 \right)
\label{3.15}
\end{equation}
In this oscillatory regime,
the Bessel function $J_n (x)$ has only a very limited dependence on $n$, namely it has
the form (\ref{3.15}) for some $n$, plus the three possible phase shifts of $\pi/2$.
This means that conditions along the chain do not vary very much for reasonably
large sections, which relates to the establishment of local equilibrium.

These classical solutions may be used to determine the time evolution
of the correlation functions such as $ \sigma (q_n, q_m) $, $ \sigma (q_n, p_m)$
and $ \sigma (p_n, p_m)$ which are the key quantities determining
the behaviour of the local densities under time evolution.
In brief, what is found is the following.
An initially uncorrelated state will develop correlations, but these
then decay with time, with the correlations never becoming too great. Furthermore,
the quantities $ (\Delta q_n)^2 $ and $ (\Delta p_n)^2 $ become dependent only very
weakly on $n$, indicating a situation similar to local equilibrium.

Now consider the local densities of the oscillator chain. For simplicity, we focus on the
number density $n(k)$, given by the one-dimensional version Eq.(\ref{6}).
Following the general scheme outlined in the previous section,
we consider initial states which are approximate eigenstates of the local
densities. Gaussian states suffice, in fact, and
these will be approximate eigenstates of the local densities if we
choose the correlation functions $ \s (q_n, q_m) $, $\s (q_n, p_m)
$ and $ \s (p_n, p_m) $ to be zero, or at least sufficiently
small, for $n \ne m $.

In a general Gaussian state, we have
\begin{equation}
\la n(k) \ra = \sum_{j=1}^N \la e^{i k q_j } \ra
= \sum_{j=1}^N \exp \left( { i k \la q_j \ra - \half k^2 (\Delta q_j)^2 } \right)
\end{equation}
and
\begin{eqnarray}
( \Delta n ( k ) )^2 &=& \la n^{\dag} (k) n(k) \ra - | \la n(k) \ra |^2
\nonumber \\
&=& \sum_{j=1}^N \sum_{n=1}^N \la e^{ i k
q_j } \ra \la e^{ - ik q_{n} } \ra \left( e^{ k^2 \s (q_j,
q_{n} )} -1 \right)
\end{eqnarray}
The latter is to be compared with
\begin{equation}
| \la n (k) \ra |^2 = \sum_{j=1}^N \sum_{n=1}^N \la e^{i k q_j } \ra
\la e^{ - i k q_{n} } \ra
\end{equation}
With an initially uncorrelated state we have $ \s (q_j, q_{n} ) = 0 $ for $ j \ne n$
and we see that
\begin{equation}
( \Delta n ( k ) )^2 = \sum_j  | \la e^{ i k q_j } \ra |^2 \left( e^{ k^2 (\Delta q_j)^2 } -1 \right)
\end{equation}
From this we expect that
\begin{equation}
( \Delta n ( k ) )^2  \ll | \la n (k) \ra |^2
\label{5.11}
\end{equation}
as long as $k^{-1}$ does not probe on scales that are too short
(compared to $ \Delta q_j$), and in this case
the Gaussian state is an approximate eigenstate as required.

Under time evolution, correlations develop, but we expect that the state will
remain an approximate eigenstate if  $k^{-1}$ is much greater than
the lengthscale of correlation. As $k$ increases from zero we have, to leading
order in small $k$,
\begin{equation}
{ ( \Delta n ( k ) )^2 \over | \la n (k) \ra |^2 } \sim { k^2 (
\Delta X)^2 \over  N^2 }
\end{equation}
where $ X = \sum_j q_j$ (the centre of mass coordinate of the whole chain).
This will be very small as long as $k^{-1}$ is much larger than
the typical lengthscale of a single particle. $ ( \Delta n ( k ) )^2 $
starts to grow very rapidly with $k$, and Eq.(\ref{5.11}) is no longer valid,
when $k^{-1}$ becomes less than the correlation length indicated
by $\s (q_j, q_{n} ) $. Hence the local density eigenstate state remains strongly peaked
about the mean as long as the coarse graining lengthscale $k^{-1}$
remains much greater than the correlation length, confirming the general
arguments of the previous section.
Similarly, it can be argued that the local density eigenstates also remain
localized in the local energy and momentum. This shows that there is
approximate decoherence of histories in the oscillator chain model,
confirming the general argument.

\section{The Approach to Local Equilibrium}

Given decoherence, we may now look at the probabilities for histories
and see if they are peaked around interesting evolution equations.
Since we have shown that there is negligible interference between
histories with an initial state consisting of a superposition of local
density eigenstates, we may take the initial state in these
probabilities to be a local density eigenstate. Decoherence alone is
not enough to get the hydrodynamic equations. Decoherence ensures that the probabilities
for histories are well-defined but the probabilities may not be peaked around
any particularly interesting histories and in fact will typically describe a situation which is highly
stochastic.
The hydrodynamic equations we seek form a {\it closed} set of equations.
This requires at least two things in the histories description.
First of all, it requires that we consider histories specified by
a sufficiently large number of variables -- all three of the local
densities, particle number, momentum, energy, are required. It is not
enough to consider histories of just one of them. Even classically,
the momentum density, for example, will generally not obey a closed
evolution equation on its own. Hence, we will assume that histories
of all three local densities are considered.

Secondly, the hydrodynamic equations emerge only when the initial state is a local equilibrium
state. We need to show how this state, a mixed state,
arises from the local density eigenstate, a pure state defined very differently.
The point here is that for sufficiently coarse grained projections onto the
local densities, the object that will determine the probabilities for histories
is $\rho_1$, the one-particle density operator constructed by tracing
the local density eigenstate. This is actually quite similar to the local equilibrium
state, since they are both mixed states localized in the local densities.
They differ in that $\rho_1$ may still contain correlations
(and in particular have non-zero $\sigma (p,q)$) not contained in the
local equilibrium state. However, since they are so similar,
it is physically extremely plausible
that $\rho_1$ will approach the local equilibrium form on short time
scales and this has indeed been explicitly verified in the oscillator model
of Ref.\cite{Hal10}. It then follows that the probabilities will be peaked about
the hydrodynamic equations.

The final picture we have is as follows. We can imagine an initial
state for the system which contains superpositions of
macroscopically very distinct states. Decoherence of histories
indicates that these states may be treated separately and we thus
obtain a set of trajectories which may be regarded as exclusive
alternatives each occurring with some probability. Those
probabilities are peaked about the average values of the local
densities. We have argued that each local density eigenstate may
then tend to local equilibrium, and a set of hydrodynamic
equations for the average values of the local densities then follow.
We thus obtain a statistical ensemble of trajectories, each of
which obeys hydrodynamic equations. These equations could be very
different from one trajectory to the next, having, for example,
significantly different values of temperature. In the most
general case they could even be in different phases, for example
one a gas, one a liquid.

Decoherence requires the coarse-graining scale
$k^{-1}$ to be much greater than the correlation
length of the local density eigenstates, and the derivation of the
hydrodynamic equations requires $k^{-2} \gg ( \Delta q)^2$. In
brief, the emergence of the classical domain occurs on
lengthscales much greater than any of the scales set by the
microscopic dynamics.

\section{Connections with Environmentally Induced Decoherence}

As noted in the Introduction, most studies of decoherence and emergent
classicality have focused on the situation in which there is an
explicit split into system and environment, and there, the decoherence
comes about due to the coarse-graining over environmental variables.
What is the connection between environmentally-induced decoherence (EID) and
conservation-induced decoherence (CID) considered in this paper? Here we
consider three different issues.

First, in EID
there is the question
of the split into system and environment. Here, the guiding princple
is conservation.  System usually means a ``large" particle, and environment
a bunch of ``small" particles, but in practice the key difference between
them is that large particles are slow and the small ones fast, which relates,
approximately, to conservation of something, such as number or momentum density.
(Although there is typically no limit of exact conservation).

Second, there is a unified way of seeing decoherence of histories in the two
cases.
Denote a generic variable by $A(t)$. Decoherence of histories of $A$ follows
when $A(t)$ commutes with itself at different times. Commutation and the resultant
decoherent and usually not exact, so approximate decoherence follows when
a condition something like this holds:
\beq
\parallel  A(t_2) A(t_1) + A (t_1) A (t_2) \parallel \  \gg \ \parallel [A(t_2), A(t_1)] \parallel
\label{uni}
\eeq
That is, the anticommutator is much bigger than the commutator in some
suitably defined operator norm $ \parallel \cdots \parallel $.

For CID, $A$ is one or more of the hydrodynamic variables
$ n(k) $, $ g(k)$, $ h(k) $. These quantities are exactly conserved at $k=0$
so commute with their values at different times at.
The inequality Eq.(\ref{uni}) can be satisfied
because the right hand side of this inequality may be made arbitrarily small
by taking $k$ sufficiently small.

For EID, $A$ is typically the position $x$ of a Brownian particle coupled
to an environment and $x(t)$ denotes evolution with the total (system plus
environment) Hamiltonian. The norm includes a trace over the environment
in a thermal state. The right hand side will be proportional to $1/M$
($M$ is the mass of the particle) which will be "small" due to the
massiveness (slowness) of the particle and it will also be proportional to $\hbar$.
However, what is more important is that, because of the thermal
fluctuations, the left hand side will be large - it is typically of
order $(\Delta x)^2$ which grows with time and with temperature of the environment.
This corresponds to the known fact that EID comes about when thermal fluctuations
are much larger than quantum ones.

In brief, Eq.(\ref{uni}) gives a unified picture of decoherence of
histories. It is satisfied in CID because the right hand side can be made
small and in EID because the left hand side can be made large.

A third issue is the question of the relative
roles of CID and EID in a given system, since one might generally
expect that both mechanisms will operate.
The point is that it is a
question of lengthscales. We have demonstrated decoherence of the
local densities starting with exact conservation at the largest
lengthscales and then moving inwards. In this way we were able to
prove decoherence without using an environment, for certain sets
of histories at very coarse-grained scales whose probabilities are
peaked about classical paths. However, in general we would like to be
able to assign probabilities to non-classical trajectories. For
example, what is the probability that a system will follow an
approximately classical trajectory at a series of times, but then
at one particular time undergoes a very large fluctuation away
from the classical trajectory? The approach adopted here, based
on conservation, would yield
an approximately zero probability for this history, to the level
of approximation used. Yet this is a valid question that we
can test experimentally. It is at this stage that an
environment becomes necessary to obtain decoherence, and indeed it
is frequently seen in particular models that when there is
decoherence of histories due to an environment, decoherence is
obtained for a very wide variety of histories, not just histories
close to classical. It is essentially a question of information.
Decoherence of histories means that information about the
histories of the system is stored somewhere \cite{GH2,Hal6}.
Classical histories need considerably less information to specify
than non-classical ones, and indeed specification of the three
local densities at any time is sufficient to specify their entire
classical histories. This is not enough for non-classical
histories, so an environment is required to store the information.

Related to this is the issue of timescales involved
in the models considered.
Decoherence through interaction with an environment involves a
timescale, which is typically exceptionally short. Here, however,
there is no timescale associated with decoherence by approximate
conservation. The eigenstates of the local densities remain
approximate eigenstates for all time. There is, however, a
timescale involved in obtaining the hydrodynamic equations,
namely, the time required for a local density eigenstate to
approach local equilibrium. In this model, this timescale is of
order $ ( \gamma \Omega )^{-1}$ (for the infinite chain in the $K
\ne 0 $ case).

Finally, and somewhat straying from the issue of EID, we briefly
comment on relations to the Boltzmann equation in these models.
It would also be of particular interest to look at CID
models involving a gas. Many-body field
theory may be the appropriate medium in which to investigate this,
following the lead of Ref.\cite{DoHa}. The decoherent histories analysis
might confer some
interesting advantages over conventional treatments.
For example, one-particle dynamics of a
gas is described by a Boltzmann equation. One of the assumptions
involved in the derivation of the Boltzmann equation is that the
initial state of the system contains no correlations, which is
clearly very restrictive \cite{Hua}.
However, in the general approach used here
it is natural to break up any
arbitrary initial state into a superposition of local density
eigenstates, and that these may then be treated separately because
of decoherence. The local density eigenstates
typically have small or zero correlations. Hence, decoherence
gives some justification for one of the rather restrictitve
assumptions of the Boltzmann equation.

\section{Summary and Discussion}

Physics would be impossible without conservation laws.
They are respected by both classical and quantum mechanics
and are the key to understanding the emergence of classical
behaviour from an underlying quantum theory.
We have outlined the process whereby local densities become effectively
classical, using local conservation as the guiding principle.
The key idea is to split the initial state
into local density eigenstates and show that they are preserved in
form under time evolution. The subsequent probabilities for
histories are peaked about the average values of the local
densities, and the equations of motion for them form a closed set
of hydrodynamic form on sufficiently large scales, provided, in
general, that sufficient time has elapsed for the local density
eigenstates to settle down to local equilibrium.

Since this account of emergent classicality is so firmly anchored
in conservation laws, and since conservation laws are so central
to physics, it seems likely that this account is very general, and
will apply to a wide variety of Hamiltonians and initial states:
as long as there are conserved quantities there is a regime nearby
of almost conserved quantities behaving quasiclassically.

An important general question is whether the quasiclassical domain derived
in this way is unique. The familiar quasiclassical domain is characterized
by local densities obeying closed sets of deterministic evolution equations.
This domain may also be referred to as
a {\it reduced description} of the quantum system, in which, at sufficiently
coarse grained scales, certain predictions are possible using only a limited set of variables,
the local densities, without having to solve the full quantum theory.
Could there be an utterly different domain, characterized by completely
different variables, but still obeying deterministic evolution equations?
That is, is there another completely different reduced description of the system?
The derivation described here rests entirely on conservation laws, and from that
point of view, the existence of another quasiclassical domain seems most unlikely,
unless there are conservation laws that we have not yet discovered. So perhaps the
appropriate question is to ask whether different reduced descriptions of the
system are possible that do not rely on conservation laws. Little is known about
this issue at present.

\section{Acknowledgements}

I am very grateful to Jim Hartle for very many discussions
on emergent classicality over a long period. I would also like
to thank Simon Saunders for encouraging me to write this article
and for many useful discussions.
%\section{Notes}

%What happens to spreading for a non-conserved quantity? E.g.
%moment in the bound case.

%Don't quite get diffusion equation
%\input refs.tex

\bibliography{apssamp}

\end{document}